\documentclass[twoside]{dis07}
\usepackage[latin1]{inputenc}
\usepackage[dvips]{graphicx,epsfig,color}
\usepackage{wrapfig,rotating}
\usepackage{amssymb,amsmath,array}

\def\MSbar{\overline{\scriptsize{\rm MS}}}
\def\DRbar{\overline{\scriptsize{\rm DR}}}
\def\as{\alpha_s}
\def\aew{\alpha_{\rm ew}}

\begin{document}
\title{Hadronic Final States}

\author{Duncan Brown$^1$, Daniel Traynor$^2$, Alexander Savin$^3$ and
  Giulia Zanderighi$^4$
\vspace{.3cm}\\
1- University of Virginia - Dept. of Physics \\
382 McCormick Rd, P.O.Box 400714, Charlottesville, VA 22904-4714, USA
\vspace{.1cm}\\
2- Queen Mary, University of London \\
Mile End, London, E1 4NS
\vspace{.1cm}\\
3- University of Wisconsin-Madison, \\
1150 University Ave., Madison, WI 5370-1390, USA
\vspace{.1cm}\\
4- CERN - Theory Division, Geneve 23, CH-1211.
}

\maketitle

\begin{abstract}
  In this summary we give a concise overview of the experimental and
  theoretical results, which were presented during the QCD and
  Hadronic Final State Working Group sessions at the DIS 2007
  workshop.
\end{abstract}

\section{Introduction}

Final states provide a unique source of information to study
elementary particle interactions. They make it possible to test our
general understanding of QCD, explore QCD evolution in limiting
regimes and also provide a solid reference for new physics searches.
It is beyond question that precise input from the Standard Model (SM)
is needed for many new physics searches that could be carried out at
the upcoming LHC, most notably precise values of $\alpha_s$ and of the
parton density functions (PDF) are needed. The role of HERA and the
Tevatron in this respect is indisputable.
In this document we summarize the current experimental and theoretical
efforts in this direction presented at our Working Group~\cite{url}.

\section{Theoretical progress}

\subsection{Higher orders and formal developments}
One of the highlights of our Working Group was the presentation of the
full NNLO calculation of $e^+e^-\to 3$ jets by Aude
Gehrmann~\cite{Gehrmann}. The main motivation for this calculation is
that the current error on $\alpha_s(M_Z) = 0.121 \pm 0.001 {\rm (exp)}
\pm 0.005 {\rm (th)}$ is dominated by theory and that this uncertainty
largely comes from missing higher orders. A reduction of this error
can be achieved by computing NNLO 3-jet event shapes and jet rates.
This means computing 2-loop matrix elements with 3 partons in the
final state, one-loop matrix elements with 4 partons and tree-level
five parton final states. Additionally one needs to develop a
formalism to cancel divergences before clustering partons into jets,
this is done with a NNLO antenna
subtraction~\cite{Gehrmann-DeRidder:2005cm}. After implementing
results in an extended version of the program EERAD2, one can obtain a
thrust distribution at NNLO for all possible color structures. The
results presented show that large cancellations between the various
color configurations occur, high precision results are therefore
needed. This work is currently in progress, together with an
implementation of other event shapes and 3-jet rates.  This will allow
a first NNLO determination of $\alpha_s$ from event shape data.

\vspace{0.5cm}

An essential ingredient in performing fixed order calculations within
any quantum field theory is a method to regulate the ultraviolet (UV)
divergences appearing in perturbative calculations. One of the most
widely used schemes is dimensional regularization (DREG), where one
extends the number of dimensions to $D = 4 - 2 \epsilon$. It has long
been known however that dimensional regularization breaks
supersymmetry (SUSY), one of the promising new physics scenarios which
might be discovered at the LHC, because the number of degrees of
freedom for fermions and bosons is different.
Precision calculations within SUSY are most conveniently done in a
scheme which preserves the boson-fermion symmetry.
An alternative to DREG is provided by dimensional reduction (DRED),
where one splits the four-dimensional gluon field in a $D$-dimensional
component and a so called $\epsilon$-scalar, a scalar whose coupling
to fermions vanishes in four dimensions ({\it evanescent coupling}).

One attractive feature of SUSY is the unification of couplings close
to the grand unification (GUT) scale.  If one wants to test
unification within SUSY one needs to relate a five-flavour $\MSbar$
coupling at the electroweak scale, $\alpha_s^{(5), \MSbar}(M_Z)$,
where all heavy degrees have been integrated out (decoupled), to a
full $\DRbar$ coupling at the GUT scale, $\alpha_s^{(\rm full),
  \DRbar}(M_{\rm GUT})$.
To obtain a next-to-next-to-leading logarithmic (NNLL) result one
needs three-loop QCD running of the $\MSbar$ coupling from $M_Z$ up to
$M_{\rm SUSY}$, the typical mass scale of SUSY particles, two-loop
decoupling and three-loop running in SUSY from $M_{\rm SUSY}$ up to
$M_{\rm GUT}$.  Robert Harlander reported on this involved
calculation~\cite{Harlander}.  The result turns out to be almost
independent of the decoupling scale.  Such a feature was also present
in the Supersymmetry Parameter Analysis (SPA 05) prescription which
was however based only on one-loop decoupling and one-loop
running.\footnote{This is not consistent, one-loop matching needs a
  two-loop running.} The difference between the two results is larger
than the experimental uncertainty by almost a factor of four.
A peculiar feature of the presented results is that the
$\MSbar$--$\DRbar$ conversion and the $\DRbar$ running both depend on
the unphysical coupling of the $\epsilon$-scalars to quarks.

\vspace{.5cm}

Higher order calculations are crucial in a variety of different
contexts. However, at higher orders one has to deal with highly
complex structures.
Johannes Bluemlein reported on a method to simplify considerably
physical quantities appearing in Quantum Field Theories (QFT) which
depend on a single scale~\cite{Bluemlein}. Single scale problems in
massless or massive perturbative calculations in QFT can be expressed
in terms of finite harmonic sums. These sums occur in the $\epsilon$
expansion of the integrals for higher order corrections to QCD
spitting functions and Wilson coefficients for space-like and
time-like processes and high-energy scattering processes such as
Bhabba scattering and many others.

The idea is to perform a Mellin transform and exploit the symmetry of
Feynman parameter integrals in Mellin space to perform
simplifications.  One derives algebraic and structural relations
between multiple harmonic sums, and uses those relations to compactify
considerably the result of higher order corrections.  Applications of
the method include the representation of ${\cal O}(\as^2)$ Drell-Yan
and Higgs-Boson production cross-sections, anomalous dimensions and
Wilson coefficients at ${\cal O}(\as^3)$, asymptotic heavy flavor
Wilson coefficients up to ${\cal O}(\as^3)$, and soft and virtual
corrections to Bhabba scattering. Whether the method can be extended
to multi-scale problems is something to explore in the future.

\vspace{.5cm}

At partonic level, a typical final state in small-x deep inelastic
scattering off nuclei and hard proton-nucleus collisions can be
characterized by the multiplicity of color-excited nucleons. Within
the Reggeon field theory, each color-exited nucleon is associated with
the unitarity cut of the pomeron exchanged between projectile and
nucleus.
The pre-QCD Abramovsky, Gribov and Kancheli unitarity cutting rules
formulated in the 1970s relate precisely these multi-pomeron exchange
contributions to total, diffractive and inelastic cross sections.

Koyla Nikolaev explained how, starting with an exact
$k_t$-factorization for hard pQCD in nuclear environment, a dramatic
revision of the AGK rules within QCD is found~\cite{Nikolaev}.
Two kinds of unitarity cut pomerons emerge which describe the color
excitation and color rotation inelastic interactions.  The departure
of the unitarity cutting rules from the ones suggested by AGK stems
from the coupled-channel features of non-Abelian intranuclear pQCD.
In the Reggeon field theory language, the results entail a large
variety of multi-pomeron couplings which vary from one universality
class for hard pQCD processes to another.

\subsection{Jets, event shapes and unintegrated PDFs}
Jets enter a variety of measurements at colliders. What a jet roughly
is is quite intuitive: a bunch of particles created in a hard
interaction moving around a common direction and depositing energy on
their way.  Theoretically, jets are tools to create a link between the
hadrons measured in detectors and perturbative calculations. Given a
set of four-vectors, they are fully specified by the choice of a
jet-clustering algorithm, its parameters and a recombination scheme. A
variety of algorithms exist (the most simple classifications are cone
types and sequential recombination type algorithms), some better than
others as far as infrared safety, sensitivity to higher order or to
hadronization effects are concerned. Once jets are defined, a number
of observables can be studied.

\vspace{.5cm}

One interesting variable is the azimuthal correlation between two jets
in QCD hard processes, $\Delta \phi \equiv |\phi_{j1} - \phi_{j2}|$.
In the presence of only two hard partons, without additional
radiation, the two jets are back-to-back, $\Delta \phi = \pi$, the
non-zero azimuthal correlation provides therefore a measurement of the
additional QCD radiation. The azimuthal correlation is sensitive to a
variety of physical effects:
it depends on the jet algorithm and recombination scheme; it is
sensitive to soft/collinear gluon radiation and non-perturbative
effects; according to the recombination procedure in the clustering it
can be affected by so-called non-global
logarithms~\cite{Dasgupta:2001sh} and it is one of the observables
commonly used for Monte Carlo (MC) tunings.  Non-global logarithms
have been the subject of many studies recently.  The current status is
that they can be resummed numerically in the large $N_c$ limit with
single-log accuracy. Some time ago, it has been shown that the use of
a jet-algorithm reduces the presence of undesired non-global
logs~\cite{Appleby:2002ke}, however recently it has been shown that
additional logs can originate from using a
jet-clustering~\cite{Delenda:2006nf}.
For the azimuthal correlation the recombination scheme turns out be to
decisive to establish the presence of non-global logs: if one uses a
recombination scheme which adds four-momenta, as is done at the
Tevatron, the observable is non-global, if on the contrary one uses an
$E_t$ weighted average azimuth, as was done at HERA, then the
observable is continuously global.

At last year's DIS workshop experimental results were presented by
Magnus Hansson which did not agree with NLO 3-jet
predictions~\cite{Hansson:2006ht}. Triggered by this lack of agreement
between data and theory theoretical work started to improve on the
theoretical accuracy.
The disagreement between fixed order perturbative calculations and
data is to be expected in the back-to-back region since in this region
multiple soft-collinear effects become important.
Yazid Delenda reported on the resummation of soft-collinear logarithms
in the azimuthal correlation distribution in the region $\pi-\Delta
\phi \sim 0$~\cite{Delenda}. Work in progress is now the matching with
fixed NLO predictions, the inclusion of power corrections and the
extension of this study to the Tevatron for which similar measurements
exist.

\vspace{0.5cm}

Event-shapes provide a continuous measure of energy and momentum of
the final state. Usually, they are computed perturbatively at NLO and
they have large soft-collinear logarithms which can be resummed at
next-to-leading logarithmic (NLL) accuracy. Additionally, they have
large hadronization corrections which give rise to $1/Q$ corrections,
where $Q$ is the hard scale of the process. These corrections can be
modeled in terms of one universal parameter $\alpha_0$. Universality
implies that if one performs a combined fit of $\alpha_s$ and
$\alpha_0$ then the values obtained should be independent of the
event-shape used in the fit. This picture proved to work well both in
$e^+e^-$ and DIS. However, it has been tested only in three-jet event
shapes, those, like the thrust, whose first non zero contribution
comes from events with three hard partons (one incoming and two
outgoing in DIS).

Andrea Banfi presented results for four-jet event shapes, whose first
non-zero contribution is due to configurations with at least four
partons in the event~\cite{Banfi}. Specifically he considered $K_{\rm
  out}$, roughly the radiation out of the plane spanned by the three
hardest partons and the $D$-parameter, proportional to the determinant
of the energy-momentum tensor, $\theta^{\alpha, \beta} \equiv \sum_h
p_h^\alpha p_h^\beta/(Q |\vec p_h|)$.  Andrea presented fits of
$\alpha_s$ and $\alpha_0$ from distributions of the $D$-parameter for
various values of $y_{\rm cut}$, the cut used to select events
containing three hard jets. It turns out that fits from the
$D$-parameter give values of $\alpha_s$ and $\alpha_0$ which are
compatible with other three-jet event-shapes, where the fit region is
limited to the region to the right of the peak of the distribution.
For $K_{\rm out}$ on the other hand the shape of the distribution did
not allow to perform a consistent fit. A possible explanation is that
the $K_{\rm out}$ distribution is sensitive to power corrections from
the four-jet region currently not included.

\vspace{0.5cm}

Final state predictions depend critically on which initial state
partons enter the hard interaction and on the energy distribution of
those partons. This is described by parton density functions. Standard
PDFs depend on the factorization scale $\mu$ and on the longitudinal
momentum fraction $x$, but they are integrated over the transverse
$k_\perp$ component. Unintegrated parton densities (uPDFs) ${\cal
  A}(x, k_\perp^2, \mu)$ keep the information over the transverse
component during the perturbative evolution. Cross sections are then
given by the convolution of uPDFs with off-shell partonic
cross-sections $\hat \sigma(x_i, k_\perp^2)$.
If one considers e.g. the cross section for a single heavy flavoured
parton as a function of $p_t$, because of momentum conservation at LO
$p_t=0$, and non zero contributions appear first at NLO. If one
instead uses uPDFs, the first non-zero contribution opens up already
at LO.  The advantage of uPDFs is therefore that by having the correct
kinematics at LO one reduces considerably the size of the NLO
corrections.
At high energies the gluon density is dominant, therefore, as a
first approximation, one can consider the contribution from gluons
uPDFs only. 

The uPDF is determined by a convolution of a non-perturbative starting
distribution ${\cal A}_0(x, \mu_0)$ and the CCFM
evolution~\cite{Jung:2000hk} $\tilde {\cal A}(x, k_\perp, \bar q)$
\begin{equation}
x {\cal A}(x, k_\perp, \bar q) = \int dx' {\cal A}_0(x') \frac{x}{x'}
\tilde {\cal A}(x, k_\perp, \bar q)\,,
\end{equation}
where the distribution ${\cal A}_0$ is parametrized at a starting
scale $\mu_0$ by 
\begin{equation}
x {\cal A}_0(x,\mu_0) = N x^{-B_g} (1-x)^4\,. 
\end{equation}
Hannes Jung presented results of fits of $B_g$ from inclusive $F_2$
and from semi-inclusive $F_{2}^c$, giving values of $B_g=0.028 \pm
0.003$ and $B_g=0.286\pm 0.002$ respectively (for
$\mu_0=1.2$GeV)~\cite{Jung}. The different value of $B_g$ from $F_c^c$
changes the uPDF significantly, since the discrepancy is not covered
by the experimental uncertainty.  To resolve the discrepancy one can
perform a fit to DIS dijet cross sections. The result for $B_g$ turns
out to be very similar to the one obtained from $F_2$. Regarding the
residual discrepancy with the determination of the $x$ component of
${\cal A}_0$ from $F_2^c$ one can observe that the rise of the gluon
distribution from $F_2^c$ comes from lowest $x$ points. If one
considers only $x> 2 \cdot 10 ^{-4}$ and uses the $F_2$ value of $B_g$
in the $F_2^c$ distribution one obtains a $\chi^2/{\rm d.o.f.} \sim
1.1$.
Finally, Hannes presented the first measurement of the intrinsic $k_t$
component of $x {\cal A}(x,\mu_0)$ assuming a Gaussian distribution
$\sim {\rm exp}(-(k_{t0}-\mu)^2/\sigma^2)$. The fit to the intrinsic
$k_t$ distribution, presented at this workshop for the first time,
gives $\sigma \sim 1.5$ and $\mu \sim 1.5$ and is consistent with a
Gaussian distribution, although other distributions are not excluded.
One additional remark about uPDFs is that one can describe the $\Delta
\Phi$ distribution between jets with Cascade using the new fit for the
uPDFs.

\vspace{0.5cm}

While NLO predictions for inclusive jet spectra typically have errors
$\sim 10-20\%$, NLO predictions for heavy jets\footnote{Experimentaly
  a heavy-jet is any jet containing at least one heavy tagged object.}
fare much worse, the uncertainty being around $\sim 40-60\%$. The NLO
prediction for heavy-jets is thus a poorly constrained NLO
calculation. Additionally, the experimental errors of the measurement
at the Tevatron are smaller than the corresponding theoretical
uncertainty.
To understand the reason for the large theoretical uncertainty it is
useful to examine the flavour production mechanisms.  At LO the only
production mechanism for heavy flavour is flavour creation (FC),
$ll\to hh$, where $l$ is a generic light flavour and $h$ denotes the
heavy flavour under study. At NLO on the other hand two other
processes enter, flavour excitation (FEX), $lh \to lh$, where the
incoming heavy quark stems from a collinear splitting of an incoming
gluon, and gluon splitting (GSP) $ll\to l(l\to hh)$. Since the NLO
processes are enhanced by collinear and soft logarithms, $\ln(p_{\rm
  t, jet}/m_Q)$ (where $m_Q$ denotes the mass of the heavy quark),
they turn out to be more important that the LO contribution. This
means that a NLO calculation, which treats FC at NLO, but where FEX
and GSP enter only at LO has very large K-factors ($\sim 5$ at the
Tevatron and $\sim 8$ at the LHC, for accessible $p_t$ values) and
therefore also very large scale dependencies.

Giulia Zanderighi showed that if one instead clusters events using an
infrared-safe, flavour algorithm, which takes into account the
difference between quark and gluon QCD production, then flavour
becomes an infrared safe quantity. This implies that no logarithms
$\ln(p_{\rm t, jet}/m_Q)$ are present and one can accordingly perform
a light-quark calculation e.g. with NLOJET++~\cite{NLOJET}. This gains
roughly a factor 3 in accuracy: K factors become $\sim 1.3$ and scale
variation uncertainties are around $10-20\%$, as is to be expected
from a true NLO calculation~\cite{Zanderighi}. The method however
requires a good understanding of single and double $b$-tagging
efficiencies, and further experimental investigation in this direction
is needed.

\subsection{Parton showers and matrix element event generators}
QCD Monte Carlo (MC) generators and simulations are vital for physics
at the LHC. It is therefore crucial to examine critically different MC
components, including perturbative aspects (parton showers, matrix
element corrections, matching), non-perturbative aspects
(hadronization, underlying event) and tuning of the event generators.
Parton showers (PS) reflect our understanding of pQCD to all-orders.
They are commonly believed to capture at least the leading logarithmic
structure. Observables sensitive to radiation in a limited phase space
region (energy flow distributions, event shapes, \dots) are sensitive
to non-global logarithms even at leading order. Is it therefore
important to re-examine the accuracy of showers in these instances in
order to establish if for example leading non-global logarithms are
erroneously tuned in the MC parameters.

The starting point of the study presented by Mrinal Dasgupta is the
fact that angular ordering (AO) catches the relevant part of these
non-global logarithmic corrections.  HERWIG, therefore, which is based
on angular ordering is expected to be close to the full resummation of
the non-global logarithms (known only in the large $N_c$ limit).
PYTHIA (before v.6.4) uses as an ordering variable, the virtuality
$m^2$ and rejects a posteriori configurations which do not respect AO,
and is expected to do worse. PYTHIA v.6.4 on the other hand, like
ARIADNE, is based on dipole phase space, and should have the full
leading logarithmic behaviour. In the study presented by Mrinal
Dasgupta these expectations are tested by comparing full resummed
results matched to NLO with predictions for parton showers for the
radiation in a slice in rapidity $\Delta \eta$ as a function of $E_t$,
the total transverse energy of hadrons in the rapidity
slice~\cite{Dasgupta}. It turns out that the above expectations are
confirmed unless one takes a quite large region of pseudorapidity,
e.g. $\Delta \eta = 3$.  For large rapidity slices a large discrepancy
is found between the new PYTHIA (v.6.4) and the resummed result.
Further studies are needed to clarify this discrepancy and to
generally understand the various parton showers in a quantitative
fashion.  In general, it has been stressed that whenever possible one
should compare results from HERWIG and PYTHIA, and when possible of
other Monte Carlos.

\vspace{.5cm}

As stressed above, Monte Carlos today are indispensable tools for many
experimental studies. One of their very conventional application is to
estimate non-perturbative (NP) effects. NP effects at hadron colliders
include both final state hadronization effects and effects due to the
underlying event (UE).
Single jet inclusive distributions, parametrized in terms of a
jet-radius $R\equiv \sqrt{(\Delta y)^2 +(\Delta \phi)^2}$ have
$1/p_{t, \mathrm{jet}}$ power corrections due to the hadronization and
UE.

Lorenzo Magnea~\cite{Magnea} explained that it is possible to show
analytically that hadronization corrections are distinguishable from
UE because they exhibit a singular dependence on $R$, while UE effects
are proportional to the radiation they collect in a given region and
so scale as $\sim \pi R^2$.
Lorenzo then presented a MC study where one runs a MC at parton level
(p), hadron level without UE (h) and with UE (u). One then selects
events with two hard jets with the hardest jet in a chosen $p_t$ range
and one defines for each hadron level a measure of the size of the
power correction
\begin{equation}
\Delta p_T^{(h/u)} \equiv \frac12 \left(
p_{T,1}^{(h/u)}
+p_{T,2}^{(h/u)}
-p_{T,1}^{(p)}
-p_{T,2}^{(p)}
\right)\,,\qquad 
\Delta p_T^{(u-h)}
\equiv 
\Delta p_T^{(u)}
-\Delta p_T^{(h)}\,.
\end{equation}
One can examine $\Delta p_t^{(h/u)}$ divided by the leading behaviour
in $R$ ($1/R$ for hadronization effects and $R^2$ for UE) for two
different partonic channels, e.g. a quark dominated channel, $qq\to
qq$, and a gluon dominated one, $gg\to gg$, both for HERWIG and
PYTHIA. With HERWIG the result of this operation is a flat
distribution as a function of $R$, signaling that the leading $R$
dependence has been correctly extracted. Additionally, one can see
that the size of the UE is independent on the hard partonic scattering
channel.
With PYTHIA on the contrary the UE seems to ``know about the hard
scattering'' i.e. instead of a flat distribution one has a leftover
$1/R$ dependence. This is perhaps surprising. Additionally the UE is
sensitive to the hard channel (increases when going from quarks to
gluons).
A correct modeling of the UE is the basis for precision phenomenology
at the LHC, since for example this input enters into the determination
of the jet energy scale. These types of discrepancies should therefore
be addressed soon.
The work presented here is based on disentangling hadronization from
UE effects by exploiting their different $R$ dependences. It was
therefore recommended that, whenever possible, measurements for
different values of the jet-radius $R$ should be provided.

\vspace{.5cm}

In a more general context parton showers rely on the universal soft
and collinear (SC) factorization of the QCD matrix elements. This is a
universal property and is true to all perturbative orders. In an ideal
world this should be the only approximation made in MCs. However,
practically all current MCs are subject to many other approximations,
e.g. interference diagrams are treated only approximately with angular
ordering, the color algebra is valid only in the large $N_c$ limit,
spin treatment is usually inexact, other first approximations are
carried in order to simplify phase space and further arbitrary
techniques are often employed.

Zoltan Nagy reported on ongoing work to formulate a parton shower
which removes all the above approximations~\cite{Nagy}. The method is
based on recursive equations that can be used to generate a lowest
order parton shower for hard scattering in hadron-hadron collisions.
The formalism is based on the factorization of soft and collinear
interactions from harder interactions. It incorporates quantum
interference between different amplitudes in cases where the
interference diagrams have leading soft or collinear singularities. It
incorporates the color and spin information for the hard partons. One
of the motivations for this more formal development of a parton shower
is to have a method that can be merged to NLO calculations in a
natural way.

\vspace{.5cm}

While parton showers are crucial for many studies, it is also
well-known that in many cases parton showers fail dramatically.  This
is for instance the case in events with many hard jets: a parton
shower has the full matrix element only for the primary hard
scattering involving few partons, while all other emissions are
treated in the soft-collinear approximation.\footnote{Traditionally,
  using the soft/collinear approximation means that there is a deficit
  of radiation in the large angle region. This is however not always
  the case, one can enhance the shower arbitrarily so as to fill, even
  overfill, those regions, but this is artificial and ad-hoc.}
Therefore, any study which looks at properties of events with many
hard well-separated particles in the final state should be based on
tools which go beyond the soft/collinear approximation of parton
showers.  Today tree level matrix element generators (ALPGEN, Helac,
MadGraph/MadEvent, Amegic++, \dots) exist which can treat according to
the process up to around 5-8 particles in the final state including
full spin correlations/interference.
Simon Visscher reported on progress in Madgraph/MadEvent
(MG/ME)~\cite{Visscher}.  Specifically, MG/ME have now two different
matching schemes, an MLM type based on event rejection after the PS
and a CKKW type based on Sudakov reweighting and vetoing on showers.
This work set a basis for a systematic comparison between different
generators, matching techniques and shower
algorithms.~\cite{Maltoni:2007tc}

As an example Simon presented results obtained with MG/ME for $t\bar
t+0,1,2,3\dots$ jet events. These processes are important for a
variety of reasons. In addition to their interest {\em per se}, they
are a background to $t\bar t H$ production and in two-Higgs Doublet
Models (2HDM) $pp \to W^+W^- b \bar b b \bar b$ could be the most
interesting channel to discover the charged Higgs. In this case one
needs a reliable $t\bar t+0,1,2,3\dots$ jet event sample. If one looks
at the rapidity distribution of the hardest jet, predictions obtained
with Pythia are depleted of radiation in the central rapidity region.
Full matrix element based predictions are not, showing the need to go
beyond SC approximation of parton showers for this type of analysis.

\vspace{0.5cm}

Despite the fact that QCD and EW corrections are nothing but higher
order corrections dictated by the same SM Lagrangian, traditionally
the two corrections have mostly been treated separately. Alessandro
Vicini reported on a calculation of combined QCD (${\cal O}(\alpha_s)
+{\rm PS}))$ and EW (${\cal O}(\alpha_{\rm ew}) +{\rm PS})$
corrections to charged current Drell-Yan (DY) processes at hadron
colliders~\cite{Vicini}.
DY processes are unique at hadron colliders.  The presence of high
$p_t$ leptons, together with the large cross section makes their
detection very easy. DY processes are useful for validating pdfs, as
luminosity monitors or to obtain precision measurement of EW
observables ($M_W, \Gamma_W...$).  $W/Z$ production in association
with jets is an important background to SM and beyond Standard Model
(BSM) signals, especially to new gauge bosons. In pure QCD the
state-of-the-art are fully differential NNLO predictions
exist~\cite{Melnikov:2006kv}.

Alessandro Vicini presented results obtained with the HORACE event
generator, which includes state of the art EW radiative corrections to
DY processes including exact ${\cal O}(\alpha_{\rm ew})$ radiative
corrections matched to multiple photon radiation via PS. It's a fully
exclusive event generator, which can be easily interfaced to QCD
showering programs like HERWIG, events are saved in a Les Houches
compliant form and it can be interfaced to the LHAPDF package.
The QCD and EW corrections are matched with an approximate additive
procedure, which works to ${\cal O}(\as\aew)$ as long as hard
non-collinear radiation effects are not important. Beyond the additive
approximation a full two-loop ${\cal O}(\as\aew)$ calculation is
needed.
Results presented for $M_T^W$ and $p_\perp^l$ show that around the
jacobian peak positive QCD corrections compensate negative EW
corrections and that EW effects are mandatory in order to extract
$M_W$, only QCD PS are not sufficient. Another interesting QCD-EW
interplay effect lies in the fact the convolution with QCD PS modifies
the relative size and shape of the EW corrections. Due to the presence
of EW logarithms, EW effects become important in the tails of
distributions ($M_T^W \sim 2$TeV, $p_\perp^\mu \sim 1$TeV) they can
amount to up to $\sim {\cal O}(30)\%$ corrections, however
cross-sections are very tiny in those regions.

\section{Particle production in $ee$, $ep$, $p\bar{p}$ and $pp$ collisions}

\subsection{Pentaquark searches}
In 2004 a comprehensive program to search for pentaquarks (PQ) in high
statistic, high resolution experiments was started by the CLAS
collaboration at Jefferson Lab. P.~Rossi (CLAS) presented recent
results for the four photoproduction channels studied so
far~\cite{rossi}. No evidence for a PQ signal was observed. From these
results CLAS set an upper limit on the PQ production cross section on
protons and neutrons.

M.~Del Degan (H1) presented the invariant mass spectrum of the $\Xi
\pi$ system, studied using deep inelastic scattering (DIS) data
collected with the H1 detector at HERA~\cite{deldegan}. A clear signal
for the well established unexotic $\Xi(1530)^{0}$ baryon is observed.
Despite having similar statistics as NA49, the $\Xi^{--/0}$ PQ signal
could not be confirmed. Upper limits for $\Xi^{--/0}$ PQ production
from H1 are in agreement with those published by the ZEUS
collaboration.

\subsection{Deuteron and antideuteron production}
A first measurement of the production of deuterons and antideuterons
in DIS was presented by S.~Chekanov (ZEUS)~\cite{chekanov1}. It is
interesting to mention that this is the first measurement of deuteron
production in elementary particles collisions. Production rates of
(anti)deuterons are significantly reduced relative to the production
of (anti)protons, consistent with other world measurements. The
production rate of deuterons was measured to be approximately three
times larger than for antideuterons. The production rates were also
studied in terms of the coalescence model.

\subsection{Exclusive final states in $e^+e^-$}
BaBar has now taken over $400fb^{-1}$ of data and has used this data
in a variety of tests of QCD using different production channels.  The
first observation of $e^+e^- \rightarrow \rho^0 \rho^0$ and $e^+ e^-
\rightarrow \phi \rho^0$ was reported by S.~Saremi
(BaBar)~\cite{saremi}. The final states for these channels are even
under charge conjugation. Such processes can be understood in terms
the Two-Virtual-Photon Annihilation model. Other results presented
were a new test of factorization in the $\overline{B}0 \rightarrow
D^{*} + \omega \pi^{-}$ channel, for which there is good agreement,
and a measurement of the $\eta$ and $\eta'$ transition form factors,
were the ratio of the form factors is inconsistent with theoretical
predictions.

\subsection{Identified particle measurements in $ep$ and $pp$}
Recent results on strange particle production ($K^\pm, K^0_S, \Lambda
's$) and Bose-Einstein correlation's (BEC) between kaons were
presented in a talk given by B.~Levchenko (ZEUS)~\cite{levchenko}.
The parameters for the BEC agree well for different kaons and are
consistent with H1 and $e^+e^-$ results. No sizeable barion-antibarion
asymmetry was observed. The ratio of baryons to mesons in the resolved
photoproduction regime is much larger than in $e^+e^-$ and is not
described by Pythia predictions.

This observation agrees well with similar results from $pp$, presented
by M.~Heinz (STAR)~\cite{heinz}. STAR also presented the $p_T$ spectra
for different particles, which for the first time can be described by
the NLO pQCD predictions due to improved fragmentation functions (FF)
for baryons and strange particles using the light-flavour separated
measurements in $e^+e^-$ collisions from OPAL.

\subsection{Fragmentation functions at HERA}
The scaled momentum spectra in the current region of the Breit frame
in DIS $ep$ scattering at HERA was measured by ZEUS, presented by
B.~Brzozowska, with high precision data covering a large range of
energy scale ($5$ to $170~\rm GeV$)~\cite{brzozowska}.  The NLO pQCD
theoretical predictions cannot reproduce the data in the entire phase
space and cannot describe the $Q^{2}$ evolution of the $x_P$
distribution.  Differences between predictions using different FF were
found to be small.

\subsection{Particle production in jets}
The relative roles of perturbative and non-perturbative QCD in the
development of jets have been studied by CDF. In his talk L.~Pinera
presented the momentum distributions and multiplicity of charged
particles in jets, momentum correlations of particles and their $k_T$
distribution~\cite{pinera}. The energy scale range covered in this
analysis goes from $20$ to $160 ~\rm GeV$. The data are generally well
described by pQCD within MLLA approach, but at large $k_T$ the theory
predictions significantly deviates from data.

\section{Particle production in collisions with heavy-ions}

\subsection{Property of the Quark-Gluon Plasma}
Hadronic final states and their correlations were used by the PHENIX
collaboration (talk presented by C.~Ogilvie) to investigate properties
of the Quark-Gluon Plasma (QGP) and it's impact on particles passing
through it~\cite{ogilvie}. Energy loss and meson suppression deliver
information on the density of QPG while direct photons are insensitive
to the QPG.  These properties can aid the comparison of collisions of
protons with that of heavy-ions. The broadening of jets can be
explained in terms of induced gluon radiation in the QGP. High $p_T$
jet production is biased towards being produced on the surface of the
QGP, the second jet in these types of events has to then travel
through the QGP to the far side. Studies of the far side reveal that
the second jet is suppressed and takes on the properties of the medium
it travels through.  Studies show that the medium also responds to the
passage of the second jet and heavy quarks are seen to lose less
energy than light quarks.

\subsection{Colour transparency}
A study of colour transparency (CT) was presented by M.~Holtrop
(CLAS)~\cite{holtrop}.  The CT was studied in terms of nuclear
transparency $T(A,Q^2)$ and is clearly observed in $\rho^0$
electroproduction, where $T(A,Q^2)$ was measured to increase with
increasing $Q^2$, in a good agreement with predictions of the
theoretical model by Kopelovich et al.

\subsection{Hadronisation in nuclear environment}
Recent results from HERMES were presented by Z.~Akopov and Y. van
Haarlem~\cite{akopov,vanhaarlem}. Detailed studies of hadronisation on
many nuclear targets were performed. Substantial nuclear attenuation
was observed as a function of different kinematic variables. For the
first time these studies were done using double-differential
distributions. For the first time the formation length of the nuclear
attenuation was studied.  Absolute measurement of the $p_T$-broadening
has been performed using different targets and hadron types for
different kinematic variables.

\section{Jet production}

\subsection{Prompt photons}
S.~Chekanov (ZEUS) presented recent results on prompt photons with an
associated jet in photoproduction at HERA~\cite{chekanov2}. Prompt
photon identification was based on conversion probabilities measured
by a dedicated detector - Barrel Pre-sampler. Both PYTHIA and HERWIG
fail to describe the shape and normalization of the cross
sections. Best agreement is achieved using the $k_T$-factorization
approach. An improved agreement with other NLO pQCD calculations was
achieved by hardening the cut on the transverse momentum of the
photon.

The H1 collaboration (talk by K.~M{\"u}ller) has performed prompt
photon studies in DIS ~\cite{mueller}. The Prompt photon signal was
extracted using a log-likelihood approach based on a shower shape
analysis. The measurements were done both for isolated photons and
photons with jets. The LO MC models underestimate the data but the
shapes are generally described. NLO pQCD calculations were not yet
available.

O.~Atramentov (D\O\ ) presented inclusive photon+jet results in
$p\bar{p}$ collisions~\cite{atramentov}. Triple differential cross
sections were measured as well as the ratios of these cross sections,
significantly reducing the effect of experimental and theoretical
correlated uncertainties. Four different photon+jet topologies were
studied to explore different kinematic regions of the gluon
distribution functions. For some kinematic regions quantitative
deviations from current QCD predictions are observed.

\subsection{Jets photoproduction at HERA}
High-$E_T$ jets in photoproduction (PHP) measured by ZEUS and
presented by H.~Perrey show sensitivity to the gluon PDF and have the
potential to further constrain the parton densities of the proton and
photon~\cite{perrey}. None of the compared photon PDFs provided an
adequate description of the ZEUS resolved data, although the direct
enriched cross sections, which are less sensitive to the photon PDF,
are in good agreement with NLO pQCD calculations.

The photoproduction of dijet events with a large rapidity gap between
the jets was used by ZEUS to estimate the colour-singlet exchange
contribution, as reported by A.~Savin (ZEUS)~\cite{savin}. Multi-jet
photoproduction was used to study the properties of Multi-Parton
Interactions (MPI) at HERA. The low-invariant-jet-mass region can only
be described by adding the MPI to the MC simulation, thus providing a
good testing ground for different MPI models.  This information is
very important for understanding the structure of underlying events at
the LHC.

\subsection{Study of underlying events in DIS}
Further studies on underlying events at HERA were performed by H1
using the DIS data by measuring jets with low transverse momenta.
S.~Osman (H1) demonstrated in his talk, that the standard QCD MC
models fail to describe the data in different azimuthal regions with
respect to the leading jet~\cite{osman}. Adding the MPI significantly
improves the description at low $Q^{2}$ where resolved photon
processes are expected to contribute, but still fails at high $Q^2$.
Due to lack of NLO pQCD calculation suitable for this analysis no
conclusion from high-order contribution was drawn.

\subsection{Charge current jets at HERA}
The first charge current (CC) $ep$ jet analysis of the HERAII data was
presented by H.~Wolfe (ZEUS) ~\cite{wolfe}. For the first time the
three- and four-jet events in CC DIS are analyzed. Inclusive jets and
multi-jet cross sections were measured. Total polarized inclusive jet
cross sections agree with predictions of the Standard Model.
  
\subsection{Jets in $ee$}
An investigation of high momentum hadron and jet production by the
OPAL collaboration was presented by
A.~Krasznahorkay~\cite{krasznahorkay}.  Previously measured exclusive
jet cross sections by L3 was not well reproduced by the pQCD NLO
predictions. The new measurement by OPAL is well described by the
theory. When compared in the same kinematic regime as L3, the OPAL and
L3 measurement do not agree well and this difference has to be
understood. The $p_T$ spectrum of hadrons is well described by the
calculation except in the very high $p_T$ bin.

\subsection{Jets in $p\bar{p}$ and $pp$}
A preliminary measurement of the $b \bar b$ dijet production cross
section and the angular correlation of the jets was presented by
S.~Vallecorsa (CDF)~\cite{vallecorsa}.  The NLO calculations describe
the data well. The $b \bar b$ angular correlation shows that events
are mainly produced by the flavour creation mechanism. The low $\Delta
\phi$ tail suggests non-negligible contribution from other
processes. It was demonstrated that the inclusion of the underlying
event in the simulation significantly improves the description of the
data.

M. D'Onofrio (CDF) reported on measurements of $W$+jets and $Z$+jets
production cross sections at CDF~\cite{donofrio}. Boson+jet production
channels are fundamental to test pQCD, underlying events and new
theoretical calculations. Comparisons to different calculations were
shown.  The size of non-perturbative corrections was also
estimated. The measurement is important since it gives an estimate of
background for new physics like top, Higgs and SUSY searches.

Jet production measurements was also reported by J.~Cammin (D\O\
)~\cite{cammin}. Good agreement with NLO pQCD predictions over a large
$p_T$ range was demonstrated. Systematic uncertainties are dominated
by the jet energy scale. Measurements of jets with a $\mu$ tag, used
to identify Heavy Flavor jets, and $Z$+jet production were also
presented.

Prospects for inclusive jet cross section measurements with early data
at ATLAS were presented by D.~Clements~\cite{clements}. Experimental
errors are expected to be dominated by the jet energy
scale. Theoretical errors at high $p_T$ are dominated by the
uncertainty on the high-x gluon PDF.

\subsection{Inclusive jet production and jet correlations in DIS}
Recent results on angular correlations in three-jet events and jet
substructure were presented by E.~Ron (ZEUS)~\cite{ron}. Fixed-order
calculations, separated according to the colour configurations, were
used to study the sensitivity of the angular distributions to the
underlying gauge structure. Using different variables one can
distinguish between contributions from triple-gluon vertexes in
quark-induced and gluon-induced processes. The measurements are
consistent with the admixture of colour configurations as predicted by
$SU(3)$ and disfavour some other combinations. Subjet structure is
reasonably described by the fixed-order QCD calculations and are
consistent with the dominance of quark-induced processes.

Inclusive jet production in DIS at high $Q^2$ is a well established
measurement and is well reproduced by different pQCD calculations.
This is why this regime is used for precise QCD measurements and
extraction of $\alpha_S$. An updated H1 analysis with improved
precision was presented by T.~Kluge (H1)~\cite{kluge}. The $\alpha_S$
fit was performed minimising the experimental uncertainty.  A first
extraction of $\alpha_S$ using the ratio of inclusive jet to the
inclusive DIS cross sections was made which provides a more precise
measurement.  The ZEUS measurements were presented by
T.~Schoerner-Sadenius~\cite{schoerner}. Together with inclusive cross
sections ZEUS measured dependence of inclusive jet cross section on
jet radius and used this measurement to extract the $\alpha_s$ value.
The inclusive dijet sample was analyzed using combined HERAI and
HERAII data, correspondent to integrated luminosity of $210$ pb$^-1$.
For all these measurements the theoretical uncertainties still
dominate.

Inclusive jet production cross section in the low $Q^2$ region, $5 <
Q^2 < 100~\rm GeV^{2}$, was presented by A.Baghdasaryan
(H1)~\cite{baghdasaryan}.  The analysis extends to lower $Q$ the
similar analysis performed by H1 at high $Q^{2}$. Good agreement with
NLO pQCD calculation was observed for $Q^{2}$ above $10~\rm GeV^{2}$.
For further talks on jet production in the low-$x$ regime see forward
jet production and HERA multi-jet production at low-$x$ and low-$Q^2$
presented during a special Structure Function WG session.

\section{Summary}

In addition to several formal theoretical developments, further
experimental and theoretical contributions demonstrated great progress
towards a more precise understanding of QCD final state production.
Enormous progress in the theoretical calculations triggered by more
precise data included a first resummation for QCD 3-jet production at
Next-to-Leading Order which should provide a better description of
dijet azimuthal correlation measurements both at HERA and the
Tevatron. An additional dedicated session was held to discuss
$\alpha_S$, where theoretical uncertainties dominate experimental
measurements at HERA.  NNLO calculations for 3-jet production in
$e^{+}e^{-}$ were presented and are expected to lead to new fits for
$\alpha_{s}$ with reduced theoretical uncertainties.

Future new physics searches at the LHC will rely on the most precise
determinations of the parton density functions.  HERA and Tevatron
photon + jet production measurements, highlight regions where
continuing improvements to theoretical models and inputs are still
required and measurements of high $E_{T}$ forward dijet production at
HERA and inclusive jet production at the Tevatron should also help
further constrain these inputs. Progress towards a precision
determination of unintegrated PDFs and a first determination of their
intrinsic $k_{T}$ component was also presented at the meeting.

Tests of power corrections arising from hadronization effects have
been extended to 3-jet event shape distributions. Progress in
perturbative evolution and hadronization issues, the role of the
underlying event implementation in Monte Carlos, reducing model
approximations for parton showers and issues that arise when Monte
Carlos are tuned to non-global parameters were highlighted. The
importance of multiple interaction modeling was demonstrated by HERA
$\gamma p$ multijet production measurements and $b \bar{b}$ jet
azimuthal correlations at the Tevatron. Improvements using a new
flavor jet algorithm to theoretical uncertainties for heavy flavor jet
production were demonstrated.

Recent exclusive final state measurements include an observation of
two virtual photon annihilation final states at BaBar. Work in
progress to resolve pentaquark issues currently show no further
evidence for pentaquark states. A first measurement of the ratio of
deuteron and antideuteron production rates in DIS at HERA is smaller
than expected in the coalescence model. NLO pQCD predictions using
improved fragmentation functions are found to describe the measured
particle $p_{t}$ spectra at RHIC, however the effect of fragmentation
functions is small and unable to account for observed differences to
NLO pQCD predictions in scaled momentum spectra measurements in DIS at
HERA. In collisions with heavy ions recent measurements at RHIC on the
impact of Quark-Gluon plasma properties to particles passing through
it were presented, together with first double differential nuclear
attenuation measurements on a series of different nuclear targets at
HERMES and color transparency effects which were reported in
$\rho^{0}$ electroproduction measurements at CLAS.  Further details
for each individually presented topic can be found in the dedicated
contributions to these proceedings.


\begin{footnotesize}

\end{footnotesize}


\end{document}